# Exact $L^2$ series solution of the Dirac-Coulomb problem for all energies

A. D. Alhaidari

*Physics Department, King Fahd University of Petroleum & Minerals, Box 5047, Dhahran 31261, Saudi Arabia*
e-mail: haidari@mailaps.org

We obtain exact solution of the Dirac equation with the Coulomb potential as an infinite series of square integrable functions. This solution is for all energies, the discrete as well as the continuous. The spinor basis elements are written in terms of the confluent hypergeometric functions and chosen such that the matrix representation of the Dirac-Coulomb operator is tridiagonal. The wave equation results in a three-term recursion relation for the expansion coefficients of the wavefunction which is solved in terms of the Meixner-Pollaczek polynomials.

## I. INTRODUCTION

Sometimes, it is argued that exact solutions of the wave equation are by some (debatable) definitions "trivial". Nevertheless, exact solutions are important because of the conceptual understanding of physics that can only be brought about by the analysis of such solutions. In fact, exactly solvable problems are valuable means for checking and improving models and numerical methods being introduced for solving complicated physical problems. Furthermore, in some limiting cases or for some special circumstances they may constitute analytic solutions of realistic problems or approximations thereof. In nonrelativistic quantum mechanics, the search for exact solutions of the wave equation was carried out over the years by many authors where several classes of these solvable potentials are accounted for and tabulated (see, for example, the references cited in [1]). Most of the known exactly solvable problems fall within distinct classes of what is referred to as "shape invariant potentials". Supersymmetric quantum mechanics [2], potential algebras [3], and point canonical transformations [4] are three methods among many which are used in the search for exact solutions of the wave equation. These developments were extended to other classes of conditionally exactly [5] and quasi exactly [6] solvable problems where all or, respectively, part of the energy spectrum is known. Recently, the relativistic extension of some of these formulations was carried out where several relativistic problems where formulated and solved exactly [7].

In all of these developments, the objective is to find solutions of the eigenvalue wave equation $H|\chi\rangle = E|\chi\rangle$, where $H$ is the Hamiltonian and $E$ is the energy which is either discrete (for bound states) or continuous (for scattering states). In most cases, especially in the search for algebraic or numerical solutions, the wave function $\chi$ is expanded in terms of discrete square integrable basis $\{\psi_n\}_{n=0}^{\infty}$ as $|\chi(\vec{r},E)\rangle = \sum_n f_n(E)|\psi_n(\vec{r})\rangle$, where $\vec{r}$ is the configuration space coordinate. The basis functions must be compatible with the domain of the Hamiltonian and should satisfy the boundary conditions. Typically the choice of basis is limited to those that carry diagonal

representations of the Hamiltonian. That is, one looks for an $L^2$ basis set $\{\psi_n\}_{n=0}^{\infty}$ such that $H|\psi_n\rangle = E_n|\psi_n\rangle$ giving the discrete spectrum $\{E_n\}$ of $H$. The continuous spectrum is obtained from the analysis of an infinite sum of these *complete* basis functions. Truncating this sum, for numerical reasons, may create problems such as the presence of unphysical states or fictitious resonances in the spectrum.

In this article we relax the restriction of a diagonal representation of the Hamiltonian by searching for square integrable bases that could support a tridiagonal matrix representation of the wave operator. That is, the action of the wave operator on the elements of the basis is allowed to take the general form $(H-E)|\psi_n\rangle \sim |\psi_n\rangle + |\psi_{n-1}\rangle + |\psi_{n+1}\rangle$ such that

$$\langle \psi_n | H - E | \psi_m \rangle = (a_n - x)\delta_{n,m} + b_n \delta_{n,m-1} + b_{n-1} \delta_{n,m+1} \tag{1.1}$$

where $x$ and the coefficients $\{a_n, b_n\}_{n=0}^{\infty}$ are real and, in general, functions of the energy $E$, the angular momentum $\ell$, and the potential parameters. Therefore, the matrix representation of the wave equation, which is obtained by expanding $|\chi\rangle$ as $\sum_m f_m |\psi_m\rangle$ in $(H-E)|\chi\rangle = 0$ and projecting on the left by $\langle \psi_n |$, results in the following three-term recursion relation

$$xf_n = a_n f_n + b_{n-1} f_{n-1} + b_n f_{n+1} \tag{1.2}$$

Consequently, the problem translates into finding solutions of the recursion relation for the expansion coefficients of the wavefunction. In most cases this recursion is solved easily and directly by correspondence with those for well known orthogonal polynomials. An example of a problem which is already solved using this approach is the non-relativistic Coulomb problem where the expansion coefficients of the wavefunction are written in terms of the Pollaczek polynomials [8]. It is obvious that the solution of (1.2) is obtained modulo an overall factor which is a function of $x$ but, otherwise, independent of $n$. The uniqueness of the solution is achieved by the requirement that the wavefunction $|\chi(\vec{r}, E)\rangle$ be energy normalizable.

It should be noted that the solution of the problem as given by Eq. (1.2) above is obtained for all $E$, the discrete as well as the continuous, constrained only by the reality and boundedness of the tridiagonal representation. Moreover, the representation equation (1.1) clearly shows that the discrete spectrum is easily obtained by diagonalization which requires that:

$$b_n = 0, \text{ and } a_n - x = 0 \tag{1.3}$$

In Sec. II, we set up the three dimensional Dirac equation for a charged spinor interacting with the electromagnetic four-potential $(A_0, \vec{A})$. Spherical symmetry is imposed and we consider the special case where the space component of the electromagnetic potential vanishes (i.e., $\vec{A} = 0$). The time component, on the other hand, is taken as the Coulomb potential. As a result, the problem is reduced to solving the radial Dirac equation. A global unitary transformation is applied to this equation to separate the variables such that the resulting second order differential equation for the radial spinor components becomes Schrödinger-like. This results in a simple and straightforward correspondence to the well-known nonrelativistic Schrödinger-Coulomb problem. The correspondence will be used in Sec. III as a guide for constructing a square integrable basis for the solution space of the Dirac-Coulomb problem. In this construction we impose the requirement that the matrix representation of the Dirac operator be tridiagonal. The result is a three-term recursion relation for the expansion coefficients of the spinor wavefunction which is solvable for all energies, the discrete as well as the continuous. The recursion relation is written in a form that makes its solution easily attainable by simple and direct comparison with that of the Meixner-Pollaczek polynomials [9]. We conclude with a short discussion in Sec. IV.

## II. FORMULATION OF THE DIRAC-COULOMB PROBLEM

Dirac equation is a relativistically covariant first order differential equation in four dimensional space-time for a spinor wavefunction $\chi$. For a free structureless particle it reads $(i\hbar\gamma^\mu\partial_\mu - mc)\chi = 0$, where $m$ is the rest mass of the particle and $c$ the speed of light. The summation convention over repeated indices is used. That is, $\gamma^\mu\partial_\mu \equiv \sum_{\mu=0}^{3}\gamma^\mu\partial_\mu = \gamma^0\partial_0 + \vec{\gamma}\cdot\vec{\partial} = \gamma^0\frac{\partial}{c\partial t} + \vec{\gamma}\cdot\vec{\nabla}$. $\{\gamma^\mu\}_{\mu=0}^{3}$ are four constant square matrices satisfying the anticommutation relation $\{\gamma^\mu,\gamma^\nu\} = \gamma^\mu\gamma^\nu + \gamma^\nu\gamma^\mu = 2\mathcal{G}^{\mu\nu}$, where $\mathcal{G}$ is the metric of Minkowski space-time which is equal to diag$(+,-,-,-)$. These are even dimensional matrices with a minimum dimension of four corresponding to spin ½ representation of the Lorentz space-time symmetry group. A four-dimensional matrix representation that satisfies this relation is chosen as follows:

$$\gamma^0 = \begin{pmatrix} I & 0 \\ 0 & -I \end{pmatrix}, \quad \vec{\gamma} = \begin{pmatrix} 0 & \vec{\sigma} \\ -\vec{\sigma} & 0 \end{pmatrix} \tag{2.1}$$

where $I$ is the 2×2 unit matrix and $\vec{\sigma}$ are the three 2×2 hermitian Pauli matrices. In the atomic units ($\hbar = m = e = 1$), the Compton wavelength $\lambdabar = \hbar/mc = 1/c$ is the relativistic parameter and the Dirac equation reads $(i\gamma^\mu\partial_\mu - \lambdabar^{-1})\chi = 0$, where $\chi$ is a four-component spinor. Next, we let the Dirac spinor be charged and coupled to the four component electromagnetic potential $A_\mu = (A_0, \vec{A})$. Gauge invariant coupling, which is accomplished by the "minimal" substitution $\partial_\mu \to \partial_\mu + i\frac{e}{\hbar c}A_\mu$, transforms the free Dirac equation to $\left[i\gamma^\mu(\partial_\mu + i\lambdabar A_\mu) - \lambdabar^{-1}\right]\chi = 0$ which, when written in details, reads as follows

$$i\lambdabar\frac{\partial}{\partial t}\chi = \left(-i\vec{\alpha}\cdot\vec{\nabla} + \lambdabar\vec{\alpha}\cdot\vec{A} + \lambdabar A_0 + \lambdabar^{-1}\beta\right)\chi \equiv \lambdabar^{-1}H\chi \tag{2.2}$$

where $H$ is the Hamiltonian in unite of $mc^2 = 1/\lambdabar^2$, $\vec{\alpha}$ and $\beta$ are the hermitian matrices

$$\vec{\alpha} = \gamma^0\vec{\gamma} = \begin{pmatrix} 0 & \vec{\sigma} \\ \vec{\sigma} & 0 \end{pmatrix}, \quad \beta = \gamma^0 = \begin{pmatrix} I & 0 \\ 0 & -I \end{pmatrix} \tag{2.3}$$

Substituting these in Eq. (2.2) gives the following matrix representation of the Dirac Hamiltonian

$$H = \begin{pmatrix} +1 + \lambdabar^2 A_0 & -\lambdabar i\vec{\sigma}\cdot\vec{\nabla} + \lambdabar^2\vec{\sigma}\cdot\vec{A} \\ -\lambdabar i\vec{\sigma}\cdot\vec{\nabla} + \lambdabar^2\vec{\sigma}\cdot\vec{A} & -1 + \lambdabar^2 A_0 \end{pmatrix} \tag{2.4}$$

Thus the eigenvalue wave equation reads $(H - \varepsilon)\chi = 0$, where $\varepsilon$ is the relativistic energy which is real, dimensionless and measured in units of $1/\lambdabar^2$.

Now, we choose $\vec{A} = 0$ and impose spherical symmetry by taking $A_0 = V(r)$. In this case, the angular variables could be separated and we can write the spinor wave-function as [10]

$$\chi = \begin{pmatrix} i[g(r)/r]\varphi_{\ell m}^j \\ [f(r)/r]\vec{\sigma}\cdot\hat{r}\varphi_{\ell m}^j \end{pmatrix} \tag{2.5}$$

where $f$ and $g$ are real radial square-integrable functions, $\hat{r}$ is the radial unit vector, and the angular wave-function for the two-component spinor is written as

$$\varphi_{\ell m}^j(\hat{r}) = \frac{1}{\sqrt{2\ell+1}} \begin{pmatrix} \sqrt{\ell \pm m + 1/2}\ Y_\ell^{m-1/2} \\ \pm\sqrt{\ell \mp m + 1/2}\ Y_\ell^{m+1/2} \end{pmatrix}, \qquad \text{for } j = \ell \pm \tfrac{1}{2} \tag{2.6}$$

$Y_\ell^{m\pm 1/2}(\hat{r})$ is the spherical harmonic function and $m$ stands for the integers in the range $-j, -j+1, \ldots, j$ and should not be confused with the mass. Spherical symmetry gives $i\vec{\sigma}\cdot(\vec{r}\times\vec{\nabla})\chi(r,\hat{r}) = -(1+\kappa)\chi(r,\hat{r})$, where $\kappa$ is the spin-orbit quantum number defined as $\kappa = \pm(j+\tfrac{1}{2}) = \pm 1, \pm 2, \ldots$ for $\ell = j \pm \tfrac{1}{2}$. Using this we obtain the following useful relations

$$(\vec{\sigma}\cdot\vec{\nabla})(\vec{\sigma}\cdot\hat{r})F(r)\varphi_{\ell m}^j(\hat{r}) = \left(\frac{dF}{dr} + \frac{1-\kappa}{r}F\right)\varphi_{\ell m}^j$$

$$(\vec{\sigma}\cdot\vec{\nabla})F(r)\varphi_{\ell m}^j(\hat{r}) = \left(\frac{dF}{dr} + \frac{1+\kappa}{r}F\right)(\vec{\sigma}\cdot\hat{r})\varphi_{\ell m}^j \tag{2.7}$$

Employing these in the wave equation $(H-\varepsilon)\chi = 0$ results in the following 2×2 matrix equation for the two radial spinor components

$$\begin{pmatrix} +1 + \lambdabar^2 V(r) - \varepsilon & \lambdabar\left(\frac{\kappa}{r} - \frac{d}{dr}\right) \\ \lambdabar\left(\frac{\kappa}{r} + \frac{d}{dr}\right) & -1 + \lambdabar^2 V(r) - \varepsilon \end{pmatrix}\begin{pmatrix} g(r) \\ f(r) \end{pmatrix} = 0 \tag{2.8}$$

Taking $V(r)$ as the Coulomb potential $Z/r$ gives

$$\begin{pmatrix} +1 + \lambdabar^2 \frac{Z}{r} - \varepsilon & \lambdabar\left(\frac{\kappa}{r} - \frac{d}{dr}\right) \\ \lambdabar\left(\frac{\kappa}{r} + \frac{d}{dr}\right) & -1 + \lambdabar^2 \frac{Z}{r} - \varepsilon \end{pmatrix}\begin{pmatrix} g \\ f \end{pmatrix} = 0 \tag{2.9}$$

where $Z$ is the charge of the particle in our chosen units ($\hbar = m = e = 1$). It should be noted that in these units the role of the fine structure constant $\alpha$ is played by the Compton wavelength $\lambdabar$. The units ($\hbar = c = 1$) where the fine structure constant $\alpha$ is used as the relativistic parameter are suitable for the electromagnetic interaction. The units that we are adopting here, where the relativistic parameter is $\lambdabar$, are suitable for dealing with a larger class of problems. The connection between these two units is in the relation $\alpha\mathcal{Z} = \lambdabar Z$, where $\mathcal{Z}$ is the usual dimensionless charge in units of $e$. Now $\lambdabar$ has the dimension of length. Therefore, in these units $Z$ has the dimension of inverse length.

Equation (2.9) results in two coupled first order differential equations for the two radial spinor components $f$ and $g$. Eliminating the lower component in favor of the upper gives a second order differential equation. This equation is not Schrödinger-like (i.e., it contains first order derivatives). To obtain a Schrödinger-like equation we proceed as follows. A global unitary transformation $U(\eta) = \exp(\tfrac{i}{2}\lambdabar\eta\sigma_2)$ is applied to the radial Dirac equation (2.9), where $\eta$ is a real constant parameter and $\sigma_2$ is the 2×2 Pauli matrix $\begin{pmatrix} 0 & -i \\ i & 0 \end{pmatrix}$. The Schrödinger-like requirement dictates that the parameter $\eta$ satisfies the constraint $\sin(\lambdabar\eta) = \lambdabar Z/\kappa$, where $-\frac{\pi}{2} \leq \lambdabar\eta \leq +\frac{\pi}{2}$ depending on the signs of $Z$ and $\kappa$. Equation (2.9) is now transformed into the following

$$\begin{pmatrix} \frac{\gamma}{\kappa} - \varepsilon + 2\lambdabar^2 \frac{Z}{r} & \lambdabar\left(-\frac{Z}{\kappa} + \frac{\gamma}{r} - \frac{d}{dr}\right) \\ \lambdabar\left(-\frac{Z}{\kappa} + \frac{\gamma}{r} + \frac{d}{dr}\right) & -\frac{\gamma}{\kappa} - \varepsilon \end{pmatrix} \begin{pmatrix} \phi(r) \\ \theta(r) \end{pmatrix} = 0 \quad (2.10)$$

where $\gamma = \kappa\sqrt{1-(\lambdabar Z/\kappa)^2}$ and

$$\begin{pmatrix} \phi \\ \theta \end{pmatrix} = U\chi = \begin{pmatrix} \cos\frac{\lambdabar\eta}{2} & \sin\frac{\lambdabar\eta}{2} \\ -\sin\frac{\lambdabar\eta}{2} & \cos\frac{\lambdabar\eta}{2} \end{pmatrix} \begin{pmatrix} g \\ f \end{pmatrix} \quad (2.11)$$

It is to be noted that the angular parameter of the unitary transformation $U(\eta)$ was intentionally split as $\lambdabar\eta$ and not collected into a single angle, say $\varphi$. This is suggested by investigating the constraint $\sin(\varphi) = \lambdabar Z/\kappa$ in the nonrelativistic limit ($\lambdabar \to 0$) where we should have $\sin(\varphi) \approx \varphi = \lambdabar Z/\kappa$. It also makes it obvious that in the nonrelativistic limit the transformation becomes the identity (i.e., not needed).

Equation (2.10) gives the lower spinor component in terms of the upper as follows

$$\theta = \frac{\lambdabar}{\gamma/\kappa + \varepsilon}\left(-\frac{Z}{\kappa} + \frac{\gamma}{r} + \frac{d}{dr}\right)\phi \quad (2.12)$$

for $\varepsilon \neq -\gamma/\kappa$. Whereas, the resulting Schrödinger-like wave equation for the upper component becomes

$$\left[-\frac{d^2}{dr^2} + \frac{\gamma(\gamma+1)}{r^2} + 2\frac{Z\varepsilon}{r} - \frac{\varepsilon^2-1}{\lambdabar^2}\right]\phi(r) = 0 \quad (2.13)$$

Comparing this equation with that of the well-known nonrelativistic Coulomb problem

$$\left[-\frac{d^2}{dr^2} + \frac{\ell(\ell+1)}{r^2} + 2\frac{Z}{r} - 2E\right]\Phi(r) = 0 \quad (2.14)$$

gives, by correspondence, the following map between the parameters of the two problems:

$$Z \to Z\varepsilon, E \to (\varepsilon^2-1)/2\lambdabar^2, \ell \to \begin{cases} \gamma \\ -\gamma-1 \end{cases} \quad (2.15)$$

The top (bottom) choice of the $\ell$ map corresponds to positive (negative) values of $\kappa$, respectively. It should be noted that the map produced by the comparison of Eq. (2.13) to Eq. (2.14) is a "correspondence" map between the parameters of the two problems and not an equality of the parameters. That is we obtain, for example, the correspondence map $\ell \to \gamma$ but not the equality $\ell = \gamma$. In fact, $\gamma$ is not an integer while, of course, $\ell$ is. Using the parameter map (2.15) in the well-known nonrelativistic energy spectrum, $E_n = -Z^2/2(\ell+n+1)^2$, gives the following relativistic spectrum for bound states

$$\varepsilon_n = \pm\left\{1+\left[\frac{\lambdabar Z}{n+\Lambda(\gamma)+1}\right]^2\right\}^{-1/2} \quad (2.16)$$

where $n = 0,1,2,...$ and either $\Lambda(\gamma) = \gamma$ or $\Lambda(\gamma) = -\gamma-1$ depending on whether $\kappa$ is positive or negative, respectively. One can easily verify that in the nonrelativistic limit ($\lambdabar \to 0, \varepsilon \to 1+\lambdabar^2 E$), the nonrelativistic spectrum is recovered. The upper radial component of the spinor wavefunction is obtained using the same parameter map (2.15) in the nonrelativistic wavefunction

$$\Phi_n(r) \sim (\lambda_n r)^{\ell+1} e^{-\lambda_n r/2} L_n^{2\ell+1}(\lambda_n r) \tag{2.17}$$

where $\lambda_n = -2Z/(n+\ell+1)$. These findings will be used in the following section as a guide to writing down the $L^2$ spinor basis that supports a tridiagonal matrix representation of the Dirac-Coulomb operator $(H-\varepsilon)$ in Eq. (2.10).

### III. TRIDIAGONAL REPRESENTATION OF THE SOLUTION SPACE

The nonrelativistic wavefunction (2.17) and the parameter map (2.15) suggest that a square integrable basis for the upper radial spinor component, which satisfies the boundary conditions, could be written as

$$\phi_n(r) = \sqrt{\frac{\omega \Gamma(n+1)}{\Gamma(n+\nu+1)}} (\omega r)^\xi e^{-\omega r/2} L_n^\nu(\omega r) \tag{3.1}$$

where $\omega$ is a positive basis scale parameter, $\xi > 0$ and $\nu > -1$. The "kinetic balance" relation (2.12) suggests that the lower component of the spinor basis is related to the upper as

$$\theta_n \sim \lambdabar \left( \frac{\mu}{2} + \frac{\zeta}{r} + \frac{d}{dr} \right) \phi_n \tag{3.2}$$

where the parameters $\mu$ and $\zeta$ are real and will be determined as we proceed. Substituting (3.1) into (3.2) and using the differential and recursion properties of the Laguerre polynomials (shown in the Appendix) we obtain

$$\theta_n(r) = \lambdabar \tau \sqrt{\frac{\omega \Gamma(n+1)}{\Gamma(n+\nu+1)}} (\omega r)^{\xi-1} e^{-\omega r/2}$$
$$\times \left[ 2\omega(\xi+\zeta-\nu) L_n^\nu(\omega r) + (\omega+\mu)(n+\nu) L_n^{\nu-1}(\omega r) + (\omega-\mu)(n+1) L_{n+1}^{\nu-1}(\omega r) \right] \tag{3.3}$$
$$= 2\lambdabar \tau \left( \frac{\mu}{2} + \frac{\zeta}{r} + \frac{d}{dr} \right) \phi_n(r)$$

where $\tau$ is another real dimensionless parameter. Square integrability requires that either (1) we impose the more stringent requirement that $\xi > 1$, or (2) choose the parameter $\zeta$ such that the sum of the Laguerre polynomials in square brackets becomes proportional to $\omega r$. The first alternative is suitable for $\kappa > 0$ since the parameter map (2.15) and the nonrelativistic wavefunction (2.17) suggest that $\xi = \gamma + 1$ as we will find out shortly. However, for $\kappa < 0$ the same parameter map gives $\xi = -\gamma$ which violates the requirement $\xi > 1$ for $\kappa = -1$ and possibly for other values of negative $\kappa$ if the electric charge $Z$ becomes large enough. Consequently, the second choice will be used for $\kappa < 0$.

#### A. Solution for $\kappa > 0$

Now for $\kappa > 0$ and with $\xi > 1$, we can simplify the expression (3.3) for $\theta_n(r)$ by eliminating the first term inside the square brackets without affecting square integrability. That is, we take $\zeta = \nu - \xi$ which results in the following expression for the lower spinor component

$$\theta_n(r) = \lambdabar\tau(\omega+\mu)\sqrt{\frac{\omega\Gamma(n+1)}{\Gamma(n+\nu+1)}}(\omega r)^{\xi-1}e^{-\omega r/2}\left[(n+\nu)L_n^{\nu-1}(\omega r) + \frac{\omega-\mu}{\omega+\mu}(n+1)L_{n+1}^{\nu-1}(\omega r)\right]$$
$$= 2\lambdabar\tau\left(\frac{\mu}{2} + \frac{\nu-\xi}{r} + \frac{d}{dr}\right)\phi_n(r) \tag{3.4}$$

In this spinor basis $\left\{\psi_n = \binom{\phi_n}{\theta_n}\right\}_{n=0}^{\infty}$, the matrix representation of the Dirac-Coulomb operator in (2.10) reads

$$\langle\psi_n|H-\varepsilon|\psi_m\rangle = \left(\frac{\gamma}{\kappa}-\varepsilon\right)\langle\phi_n|\phi_m\rangle + 2\lambdabar^2 Z\langle\phi_n|\tfrac{1}{r}|\phi_m\rangle - \left(\frac{\gamma}{\kappa}+\varepsilon-\frac{1}{\tau}\right)\langle\theta_n|\theta_m\rangle$$
$$-\lambdabar\left(\frac{Z}{\kappa}+\frac{\mu}{2}\right)\left[\langle\theta_n|\phi_m\rangle + \langle\theta_m|\phi_n\rangle\right] + \lambdabar(\gamma+\xi-\nu)\left[\langle\theta_n|\tfrac{1}{r}|\phi_m\rangle + \langle\theta_m|\tfrac{1}{r}|\phi_n\rangle\right] \tag{3.5}$$

where we have used integration by parts in writing $\langle\phi_n|\overrightarrow{\tfrac{d}{dr}}|\theta_m\rangle = -\langle\phi_n|\overleftarrow{\tfrac{d}{dr}}|\theta_m\rangle$ since the product $\phi_n(r)\theta_m(r)$ vanishes at the boundaries $r = 0$ and $r \to \infty$. Now, we require that this representation be tridiagonal. That is, $\langle\psi_n|H-\varepsilon|\psi_m\rangle = 0$ for all $|n-m| \geq 2$. For the first two terms on the right side of Eq. (3.5) to comply with this requirement we must have $2\xi = \nu+1$. Moreover, the two terms inside the last square brackets on the right side of the equation destroy the tridiagonal structure. Thus, the multiplying factor $(\gamma+\xi-\nu)$ must vanish. Consequently, the matrix representation of the Dirac-Coulomb operator for $\kappa > 0$ is tridiagonal only if $\xi = \gamma+1$ and $\nu = 2\gamma+1$. Thus, the requirement that $\xi > 1$ is preserved for all $\kappa > 0$. Therefore, for positive $\kappa$, the two components of the radial spinor basis become

$$\phi_n^+(r) = \sqrt{\frac{\omega\Gamma(n+1)}{\Gamma(n+2\gamma+2)}}(\omega r)^{\gamma+1}e^{-\omega r/2}L_n^{2\gamma+1}(\omega r) \tag{3.6a}$$

$$\theta_n^+(r) = \lambdabar\tau(\omega+\mu)\sqrt{\frac{\omega\Gamma(n+1)}{\Gamma(n+2\gamma+2)}}(\omega r)^{\gamma}e^{-\omega r/2}$$
$$\times\left[(n+2\gamma+1)L_n^{2\gamma}(\omega r) + \frac{\omega-\mu}{\omega+\mu}(n+1)L_{n+1}^{2\gamma}(\omega r)\right] \tag{3.6b}$$

Substituting these into (3.5) and using the orthogonality and recurrence relations of the Laguerre polynomials (shown in the Appendix) we obtain the following elements of the symmetric tridiagonal matrix representation of the Dirac-Coulomb operator

$$(H-\varepsilon)_{n,n} = 2(n+\gamma+1)\left[\frac{\gamma}{\kappa}-\varepsilon-\lambdabar^2\tau^2(\omega^2+\mu^2)\left(\frac{\gamma}{\kappa}+\varepsilon-\tfrac{1}{\tau}\right) - 2\lambdabar^2\tau\mu\left(\frac{Z}{\kappa}+\frac{\mu}{2}\right)\right]$$
$$+2\lambdabar^2\omega Z - 4\gamma\lambdabar^2\tau^2\mu\omega\left(\frac{\gamma}{\kappa}+\varepsilon-\tfrac{1}{\tau}\right) - 4\gamma\lambdabar^2\tau\omega\left(\frac{Z}{\kappa}+\frac{\mu}{2}\right) \tag{3.7}$$

$$(H-\varepsilon)_{n,n-1} = -\sqrt{n(n+2\gamma+1)}\left[\frac{\gamma}{\kappa}-\varepsilon+\lambdabar^2\tau^2(\omega^2-\mu^2)\left(\frac{\gamma}{\kappa}+\varepsilon-\tfrac{1}{\tau}\right) - 2\lambdabar^2\tau\mu\left(\frac{Z}{\kappa}+\frac{\mu}{2}\right)\right] \tag{3.8}$$

If we define the following quantities:
$$p = \tau^2\omega^2\left(\frac{\gamma}{\kappa}+\varepsilon-\tfrac{1}{\tau}\right), \quad q = 2\tau\omega\left(\frac{Z}{\kappa}+\frac{\mu}{2}\right) \tag{3.9}$$

Then, the matrix representation of the wave equation $(H-\varepsilon)|\chi\rangle = 0$, where $|\chi\rangle = \sum_m f_m|\psi_m\rangle$, results in the following three-term recursion relation for the expansion coefficients of the wave-function

$$2\left[(n+\gamma+1)\rho_- + \Omega\right]f_n - \rho_+\sqrt{n(n+2\gamma+1)}f_{n-1} - \rho_+\sqrt{(n+1)(n+2\gamma+2)}f_{n+1} = 0 \tag{3.10}$$

where

$$\rho_\pm = \frac{\gamma/\kappa-\varepsilon}{\lambdabar^2 p} - \frac{\mu}{\omega}\frac{q}{p} - \left(\frac{\mu}{\omega}\right)^2 \pm 1, \quad \Omega = -2\gamma\frac{\mu}{\omega} + \frac{\omega Z-\gamma q}{p} \tag{3.11}$$

Rewriting (3.10) in terms of the polynomials $P_n(\varepsilon) = \sqrt{\Gamma(n+2\gamma+2)/\Gamma(n+1)}\, f_n(\varepsilon)$, we obtain the following recursion relation

$$2\left[(n+\gamma+1)\frac{\rho_-}{\rho_+}+\frac{\Omega}{\rho_+}\right]P_n - (n+2\gamma+1)P_{n-1} - (n+1)P_{n+1} = 0 \tag{3.12}$$

We compare this with the recursion relation satisfied by the Meixner-Pollaczek polynomial $\mathsf{P}_n^\lambda(x,\varphi)$ [9] that reads

$$2\left[(n+\lambda)\cos\varphi + x\sin\varphi\right]\mathsf{P}_n^\lambda - (n+2\lambda-1)\mathsf{P}_{n-1}^\lambda - (n+1)\mathsf{P}_{n+1}^\lambda = 0 \tag{3.13}$$

where, $\lambda > 0$ and $0 < \varphi < \pi$. Thus, $\lambda = \gamma+1$, $\cos\varphi = \rho_-/\rho_+$, $x = \Omega/\sqrt{\rho_+^2 - \rho_-^2}$ and we can write

$$f_n(\varepsilon) = \sqrt{\frac{\Gamma(n+1)}{\Gamma(n+2\gamma+2)}}\, \mathsf{P}_n^{\gamma+1}\left(\frac{\Omega}{\sqrt{\rho_+^2-\rho_-^2}},\cos^{-1}\left(\frac{\rho_-}{\rho_+}\right)\right) \tag{3.14}$$

which is defined up to a multiplicative factor that depends on $\varepsilon$ but is independent of $n$. The Meixner-Pollaczek polynomial could be written in terms of the hypergeometric function as

$$\mathsf{P}_n^\lambda(x,\varphi) = \frac{\Gamma(n+2\lambda)}{\Gamma(n+1)\Gamma(2\lambda)} e^{in\varphi}\, {}_2F_1(-n,\lambda+ix;2\lambda;1-e^{-2i\varphi}) \tag{3.15}$$

The orthogonality relation associated with these polynomials is as follows

$$\int_{-\infty}^{+\infty} \rho^\lambda(x,\varphi)\mathsf{P}_n^\lambda(x,\varphi)\mathsf{P}_m^\lambda(x,\varphi)dx = \frac{\Gamma(n+2\lambda)}{\Gamma(n+1)}\delta_{nm} \tag{3.16}$$

where $\rho^\lambda(x,\varphi) = \frac{1}{2\pi}(2\sin\varphi)^{2\lambda} e^{(2\varphi-\pi)x}|\Gamma(\lambda+ix)|^2$. Therefore, the exact $L^2$ series solution of the Dirac-Coulomb problem for $\kappa > 0$ could be written as

$$\chi(r,\varepsilon) = A^{\gamma+1}(\varepsilon)\sum_{n=0}^\infty \sqrt{\frac{\Gamma(n+1)}{\Gamma(n+2\gamma+2)}}\, \mathsf{P}_n^{\gamma+1}\left(\Omega/\sqrt{\rho_+^2-\rho_-^2},\cos^{-1}(\rho_-/\rho_+)\right)\psi_n^+(r) \tag{3.17}$$

where $A^{\gamma+1}(\varepsilon) = \sqrt{\rho^{\gamma+1}(x,\varphi)(dx/d\varepsilon)}$ is a normalization factor that makes $\chi$ energy-normalizable, while the two components of the radial spinor basis element $\psi_n^+(r)$ are those given by Eqs. (3.6).

Further analysis of these solutions, such as obtaining the discrete spectrum, is tractable only if the "kinetic balance" relation (2.12) is strictly imposed on the basis elements. That is, relation (3.2) should be identical to (2.12) which requires that

$$\mu = -2Z/\kappa, \quad \tau = \tfrac{1}{2}(\varepsilon+\gamma/\kappa)^{-1} \tag{3.18}$$

for $\varepsilon \neq -\gamma/\kappa$. In this case, the tridiagonal representation of the Dirac-Coulomb operator simplifies to

$$(H-\varepsilon)_{n,n} = 2(n+\gamma+1)\left\{\frac{\gamma}{\kappa}-\varepsilon+\tfrac{1}{4}\lambdabar^2\left[\omega^2+4(Z/\kappa)^2\right]\left(\frac{\gamma}{\kappa}+\varepsilon\right)^{-1}\right\} + 2\lambdabar^2\omega Z\varepsilon\left(\frac{\gamma}{\kappa}+\varepsilon\right)^{-1} \tag{3.7'}$$

$$(H-\varepsilon)_{n,n-1} = -\sqrt{n(n+2\gamma+1)}\left\{\frac{\gamma}{\kappa}-\varepsilon-\tfrac{1}{4}\lambdabar^2\left[\omega^2-4(Z/\kappa)^2\right]\left(\frac{\gamma}{\kappa}+\varepsilon\right)^{-1}\right\} \tag{3.8'}$$

while the parameters $\rho_\pm$ and $\Omega$ in the solution (3.17) read as follows

$$\rho_\pm = 4\frac{\varepsilon^2-1}{\omega^2\lambdabar^2} \pm 1, \quad \Omega = -4(Z/\omega)\varepsilon \tag{3.19}$$

The arguments of the Meixner-Pollaczek polynomial $\mathsf{P}_n^\lambda(x,\varphi)$, in this case, read as follows:

$$\cos\varphi = \left(4\frac{\varepsilon^2-1}{\omega^2\lambdabar^2}-1\right)\Big/\left(4\frac{\varepsilon^2-1}{\omega^2\lambdabar^2}+1\right),\quad x=-\lambdabar Z\varepsilon\big/\sqrt{\varepsilon^2-1} \qquad (3.20)$$

The range $0<\varphi<\pi$ implies that the solution obtained above in (3.17) is valid for $|\varepsilon|>1$. That is, the solution (3.17) is for energies larger than the rest mass $mc^2$ which corresponds to scattering states. Solutions for $|\varepsilon|<1$ correspond to bound states. To obtain these solutions and their discrete energy spectrum, we impose the diagonalization requirement (1.3). This requirement translates, in the case of the recursion relation (3.10), into the following conditions

$$\rho_+ = 0,\quad \Omega = 2(n+\gamma+1) \qquad (3.21)$$

giving

$$\varepsilon = \varepsilon_n^\gamma = \pm\left[1+\left(\frac{\lambdabar Z}{n+\gamma+1}\right)^2\right]^{-1/2},\quad \omega=\omega_n^\gamma = \frac{-2Z\varepsilon_n^\gamma}{n+\gamma+1},\qquad n=0,1,2,\ldots \qquad (3.22)$$

which agrees with the well-known relativistic bound states energy spectrum for the Dirac-Coulomb problem when $\kappa>0$. The corresponding spinor wavefunctions are

$$\phi_n^+(r) = \sqrt{\frac{\omega_n^\gamma \Gamma(n+1)}{\Gamma(n+2\gamma+2)}}(\omega_n^\gamma r)^{\gamma+1} e^{-\omega_n^\gamma r/2} L_n^{2\gamma+1}(\omega_n^\gamma r) \qquad (3.23a)$$

$$\theta_n^+(r) = \frac{\lambdabar}{2}\frac{\omega_n^\gamma - 2Z/\kappa}{\varepsilon_n^\gamma + \gamma/\kappa}\sqrt{\frac{\omega_n^\gamma \Gamma(n+1)}{\Gamma(n+2\gamma+2)}}(\omega_n^\gamma r)^\gamma e^{-\omega_n^\gamma r/2}$$
$$\times\left[(n+2\gamma+1)L_n^{2\gamma}(\omega_n^\gamma r) + \frac{\omega_n^\gamma + 2Z/\kappa}{\omega_n^\gamma - 2Z/\kappa}(n+1)L_{n+1}^{2\gamma}(\omega_n^\gamma r)\right] \qquad (3.23b)$$

Taking the nonrelativistic limit ($\lambdabar\to 0$, $\varepsilon\to 1+\lambdabar^2 E$) in (3.19) gives $\rho_\pm = \left(8E/\omega^2\right)\pm 1$ and $\Omega = -4Z/\omega$ resulting in the three-term recursion relation for the nonrelativistic Coulomb problem which was solved by Yamani and Reinhardt [8]. In the following subsection we obtain the solution of the Dirac-Coulomb problem for $\kappa<0$.

### B. Solution for $\kappa<0$

In this case and as stated below Eq. (3.3) we choose the parameter $\zeta$ in (3.2) such that the resulting sum of the Laguerre polynomials inside the square brackets in (3.3) becomes proportional to $\omega r$. Using the properties of the Laguerre polynomials in the Appendix one can show that

$$-2\omega\nu L_n^\nu(x) + (\omega+\mu)(n+\nu)L_n^{\nu-1}(x) + (\omega-\mu)(n+1)L_{n+1}^{\nu-1}(x) = x\left[(\omega+\mu)L_n^\nu(x) - 2\omega L_n^{\nu+1}(x)\right]$$

This means that choosing $\zeta=-\xi$ in (3.3) results in a square integrable lower spinor component without the need for a stronger constraint on the real parameter $\xi$ other than $\xi>0$. Consequently, we obtain the following expression for the lower component of the spinor basis

$$\theta_n(r) = -\lambdabar\tau(\omega+\mu)\sqrt{\frac{\omega\Gamma(n+1)}{\Gamma(n+\nu+1)}}(\omega r)^\xi e^{-\omega r/2}\left[L_{n-1}^{\nu+1}(\omega r) + \frac{\omega-\mu}{\omega+\mu}L_n^{\nu+1}(\omega r)\right]$$
$$= 2\lambdabar\tau\left(\frac{\mu}{2}-\frac{\xi}{r}+\frac{d}{dr}\right)\phi_n(r) \qquad (3.24)$$

In this basis, the matrix representation of the Dirac-Coulomb operator in (2.10) reads

$$\langle\psi_n|H-\varepsilon|\psi_m\rangle = \left(\tfrac{\gamma}{\kappa}-\varepsilon\right)\langle\phi_n|\phi_m\rangle + 2\lambdabar^2 Z\langle\phi_n|\tfrac{1}{r}|\phi_m\rangle - \left(\tfrac{\gamma}{\kappa}+\varepsilon-\tfrac{1}{\tau}\right)\langle\theta_n|\theta_m\rangle$$
$$-\lambdabar\left(\tfrac{Z}{\kappa}+\tfrac{\mu}{2}\right)\left[\langle\theta_n|\phi_m\rangle+\langle\theta_m|\phi_n\rangle\right]+\lambdabar(\gamma+\xi)\left[\langle\theta_n|\tfrac{1}{r}|\phi_m\rangle+\langle\theta_m|\tfrac{1}{r}|\phi_n\rangle\right] \quad (3.25)$$

The same arguments, which were presented below Eq. (3.5), apply to this representation as well giving, however, $\xi = -\gamma$ and $\nu = -2\gamma-1$. Thus, the resulting two components of the spinor basis for $\kappa < 0$ are

$$\phi_n^-(r) = \sqrt{\tfrac{\omega\Gamma(n+1)}{\Gamma(n-2\gamma)}}(\omega r)^{-\gamma} e^{-\omega r/2} L_n^{-2\gamma-1}(\omega r) \qquad (3.26a)$$

$$\theta_n^-(r) = -\lambdabar\tau(\omega+\mu)\sqrt{\tfrac{\omega\Gamma(n+1)}{\Gamma(n-2\gamma)}}(\omega r)^{-\gamma} e^{-\omega r/2}\left[L_{n-1}^{-2\gamma}(\omega r)+\tfrac{\omega-\mu}{\omega+\mu}L_n^{-2\gamma}(\omega r)\right] \quad (3.26b)$$

Substituting these into (3.25), we obtain the following elements of the symmetric tridiagonal matrix representation of the Dirac-Coulomb operator for $\kappa < 0$

$$(H-\varepsilon)_{n,n} = 2(n-\gamma)\left[\tfrac{\gamma}{\kappa}-\varepsilon-\lambdabar^2\tau^2(\omega^2+\mu^2)\left(\tfrac{\gamma}{\kappa}+\varepsilon-\tfrac{1}{\tau}\right)-2\lambdabar^2\tau\mu\left(\tfrac{Z}{\kappa}+\tfrac{\mu}{2}\right)\right]$$
$$+2\lambdabar^2\omega Z - 4\gamma\lambdabar^2\tau^2\mu\omega\left(\tfrac{\gamma}{\kappa}+\varepsilon-\tfrac{1}{\tau}\right)-4\gamma\lambdabar^2\tau\omega\left(\tfrac{Z}{\kappa}+\tfrac{\mu}{2}\right) \qquad (3.27)$$

$$(H-\varepsilon)_{n,n-1} = -\sqrt{n(n-2\gamma-1)}\left[\tfrac{\gamma}{\kappa}-\varepsilon+\lambdabar^2\tau^2(\omega^2-\mu^2)\left(\tfrac{\gamma}{\kappa}+\varepsilon-\tfrac{1}{\tau}\right)-2\lambdabar^2\tau\mu\left(\tfrac{Z}{\kappa}+\tfrac{\mu}{2}\right)\right] \quad (3.28)$$

Comparing these with the corresponding formulas (3.7) and (3.8) for $\kappa > 0$, shows that the only difference is in the $n$-dependent factors. In these factors, and only in these factors, the replacement $\gamma \to -\gamma-1$ takes effect. Consequently, the matrix representation of the wave equation results in the following three-term recursion relation for the expansion coefficients of the wavefunction

$$2\left[(n-\gamma)\rho_- + \Omega\right]f_n - \rho_+\sqrt{n(n-2\gamma-1)}f_{n-1} - \rho_+\sqrt{(n+1)(n-2\gamma)}f_{n+1} = 0 \qquad (3.29)$$

where the parameters $\rho_\pm$ and $\Omega$ are exactly those defined in (3.11) above. Pursuing the same development carried out in the case $\kappa > 0$, we obtain the following solution for $\kappa < 0$ and for $|\varepsilon| > 1$

$$\chi(r,\varepsilon) = A^{-\gamma}(\varepsilon)\sum_{n=0}^{\infty}\sqrt{\tfrac{\Gamma(n+1)}{\Gamma(n-2\gamma)}}\,\mathsf{P}_n^{-\gamma}\left(\Omega\big/\sqrt{\rho_+^2-\rho_-^2},\cos^{-1}(\rho_-/\rho_+)\right)\psi_n^-(r) \qquad (3.30)$$

Imposing the "kinetic balance" relation among the components of the spinor basis (3.26) results in the same parameter assignments in (3.18). It also gives the following diagonalization conditions for obtaining the discrete energy spectrum:

$$\rho_+ = 0,\ \Omega = 2(n-\gamma) \qquad (3.31)$$

resulting in

$$\varepsilon = \varepsilon_n^{-\gamma-1} = \pm\left[1+\left(\tfrac{\lambdabar Z}{n-\gamma}\right)^2\right]^{-1/2},\ \omega = \omega_n^{-\gamma-1} = -\tfrac{2Z\varepsilon_n^{-\gamma-1}}{n-\gamma}, \qquad n=0,1,2,... \qquad (3.32)$$

which complement the results obtained in (3.22) above for $\kappa > 0$. The corresponding spinor wave-functions read as follows:

$$\phi_n^-(r) = \sqrt{\tfrac{\omega_n^{-\gamma-1}\Gamma(n+1)}{\Gamma(n-2\gamma)}}(\omega_n^{-\gamma-1}r)^{-\gamma} e^{-\omega_n^{-\gamma-1}r/2} L_n^{-2\gamma-1}(\omega_n^{-\gamma-1}r) \qquad (3.33a)$$

$$\theta_n^-(r) = -\tfrac{\lambdabar}{2}\tfrac{\omega_n^{-\gamma-1}-2Z/\kappa}{\varepsilon_n^{-\gamma-1}+\gamma/\kappa}\sqrt{\tfrac{\omega_n^{-\gamma-1}\Gamma(n+1)}{\Gamma(n-2\gamma)}}(\omega_n^{-\gamma-1}r)^{-\gamma} e^{-\omega_n^{-\gamma-1}r/2}$$
$$\times\left[L_{n-1}^{-2\gamma}(\omega_n^{-\gamma-1}r)+\tfrac{\omega_n^{-\gamma-1}+2Z/\kappa}{\omega_n^{-\gamma-1}-2Z/\kappa}L_n^{-2\gamma}(\omega_n^{-\gamma-1}r)\right] \qquad (3.33b)$$

## IV. DISCUSSION

We would like to conclude with some comments. First, it is worthwhile noting that the energy spectrum in (3.22) and (3.32) shows that the lowest positive energy state is $\varepsilon_+ = \left|\varepsilon_0^{-\gamma-1}\right| = \gamma/\kappa = \sqrt{1-(\hbar Z/\kappa)^2}$ for $\kappa = -1, -2, \ldots$ The highest negative energy state, on the other hand, is $\varepsilon_- = -\left|\varepsilon_0^{-\gamma-1}\right| = -\gamma/\kappa$ for $\kappa = -1, -2, \ldots$ These two are non-degenerate states, while all others are. This is so because, $\varepsilon_n^{\gamma}\big|_{\kappa>0} = \varepsilon_{n+1}^{-\gamma-1}\big|_{\kappa<0}$ for $n = 0, 1, 2, \ldots$ and for all $\kappa$. The spinor wavefunction associated with the lowest positive energy state is obtained from (3.33) as

$$\psi_0^-(r) = \sqrt{\frac{2Z/\kappa}{\Gamma(-2\gamma)}} (2Zr/\kappa)^{-\gamma} e^{-Zr/\kappa} \begin{pmatrix} 1 \\ -\hbar Z/\gamma \end{pmatrix} \qquad (4.1)$$

where, for bound states, $Z$ is negative. Obtaining the spinor wavefunction associated with the highest negative energy state is more subtle. This is due to the fact that the "kinetic balance" relation (2.12) does not hold for this state (and only this state) since $\varepsilon = -\gamma/\kappa$. One has to redo the development in subsection III.B for this case (where, $\varepsilon = -\gamma/\kappa$ and $n = 0$) without the constraint (3.18) but with arbitrary $\mu$ and $\tau$.

The second comment we want to make has to do with the type and number of solutions of the recursion relation (3.10) or (3.29). Typically, there are two solutions to such three-term recursion relation. This could be understood by noting that the orthogonal polynomials that satisfy the recursion relation are at the same time solutions of a second order differential equation. In other words, there is a correspondence between three-term recursion relations and second order differential equations for a given set of initial relations or boundary conditions, respectively. The solutions obtained above in (3.17) and (3.30) could be termed "regular solutions." These correspond to solutions of the recursion relation in terms of polynomials of the "first kind". Polynomials of the "second kind" satisfy the same recursion relation (for $n \geq 1$) but with a different initial relation (for $n = 0$). These correspond to "irregular solutions", or in a more precise term "regularized solutions," since they are regular at the origin of configuration space while behaving asymptotically as the irregular solution. In what remains we consider the recursion relation (3.10) or, equivalently (3.12). However, the same analysis could as well be carried out for the recursion relation (3.29). For a very large positive integer $N$ the recursion relation (3.12) could be rewritten as $2zP_{N+n}(z) - P_{N+n-1}(z) - P_{N+n+1}(z) = 0$. Defining $\hat{P}_n(z) \equiv P_{N+n}(z)$, we could write it as

$$2z\hat{P}_n(z) - \hat{P}_{n-1}(z) - \hat{P}_{n+1}(z) = 0 \qquad (4.2)$$

where $z = \rho_-/\rho_+ = \cos\varphi$. This is the recursion relation of the Chebyshev polynomials. For large $n$, they are oscillatory (i.e., they behave like sine's and cosine's). The two independent oscillatory solutions of (4.2), which we will designate by $P_n^\pm(z)$, differ by a phase. The origin of this phase difference could be traced back to the initial relation ($n = 0$) of the recursion (3.12). Thus, the initial relation must have two different forms. This difference propagates through the recursion to the asymptotic solutions. One of these initial relations is homogeneous and corresponds to the regular solution, which was obtained above. The other is inhomogeneous and corresponds to the regularized solution. They could be written as:

$$2\left[(\gamma+1)\frac{\rho_-}{\rho_+}+\frac{\Omega}{\rho_+}\right]P_0^+ - P_1^+ = 0 \tag{4.3a}$$

$$2\left[(\gamma+1)\frac{\rho_-}{\rho_+}+\frac{\Omega}{\rho_+}\right]P_0^- - P_1^- = W \neq 0 \tag{4.3b}$$

where $W = W(x,\varphi)$ and is related to the Wronskian of the two solutions. For scattering problems, the phase shift is obtained by the analysis of the two solutions $P_n^\pm(z)$. Such analysis is typical of algebraic scattering methods in nonrelativistic quantum mechanics. A clear example is found in the *J*-matrix method of scattering [11].

## APPENDIX: PROPERTIES OF THE LAGUERRE POLYNOMIALS

The following are useful formulas and relations satisfied by the generalized orthogonal Laguerre polynomials $L_n^\nu(x)$ that are relevant to the developments carried out in this work. They are found on most textbooks on orthogonal polynomials [12]. We list them here for ease of reference.

The differential equation:
$$\left[x\frac{d^2}{dx^2}+(\nu+1-x)\frac{d}{dx}+n\right]L_n^\nu(x) = 0 \tag{A.1}$$
where $\nu > -1$ and $n = 0,1,2,\ldots$

Expression in terms of the confluent hypergeometric function:
$$L_n^\nu(x) = \frac{\Gamma(n+\nu+1)}{\Gamma(n+1)\Gamma(\nu+1)}\,{}_1F_1(-n;\nu+1;x) \tag{A.2}$$

The three-term recursion relation:
$$xL_n^\nu = (2n+\nu+1)L_n^\nu - (n+\nu)L_{n-1}^\nu - (n+1)L_{n+1}^\nu \tag{A.3}$$

Other recurrence relations:
$$xL_n^\nu = (n+\nu)L_n^{\nu-1} - (n+1)L_{n+1}^{\nu-1} \tag{A.4}$$
$$L_n^\nu = L_n^{\nu+1} - L_{n-1}^{\nu+1} \tag{A.5}$$

Differential formula:
$$x\frac{d}{dx}L_n^\nu = nL_n^\nu - (n+\nu)L_{n-1}^\nu \tag{A.6}$$

Orthogonality relation:
$$\int_0^\infty \rho^\nu(x) L_n^\nu(x) L_m^\nu(x)\,dx = \frac{\Gamma(n+\nu+1)}{\Gamma(n+1)}\delta_{nm} \tag{A.7}$$
where $\rho^\nu(x) = x^\nu e^{-x}$.